# On-chip photon entanglement-assisted topology loading and transfer

Haoqi Zhao[1,2], Yichi Zhang[2], Isaac Nape[3], Shuang Wu[2], Yaoyang Ji[1], Chenjie Zhang[4], Yijie Shen[4,5,*], Andrew Forbes[3,†], and Liang Feng[1,2,‡]

[1]*Department of Electrical and Systems Engineering, University of Pennsylvania, Philadelphia, PA 19104, USA*
[2]*Department of Materials Science and Engineering, University of Pennsylvania, Philadelphia, PA 19104, USA*
[3]*School of Physics, University of Witwatersrand, Johannesburg 2000, South Africa*
[4]*Centre for Disruptive Photonic Technologies, School of Physical and Mathematical Sciences, Nanyang Technological University, Singapore 637371, Singapore*
[5]*School of Electrical and Electronic Engineering, Nanyang Technological University, Singapore 639798, Singapore*

Topological protection offers a robust solution to the challenges of noise and loss in physical systems. By integrating topological physics into optics, loading and encoding quantum states into topological invariants can provide resilience to information systems in the face of environmental disruptions. Here, we demonstrate on-chip loading and entanglement-assisted transfer of photon topology, where the topological structure is coherently encoded in a single-photon spin-textured quantum state, which can be transferred, through entanglement distribution, into a non-local quantum-correlated topology shared between two entangled photons. Throughout the transfer process, the topology remains protected against substantial background noise as well as isotropic and anisotropic disturbances, while quantum correlations persist. Our framework for loading and transferring topology is compatible with quantum teleportation when ancillary photons are introduced, thereby promising the development of distributed quantum systems with inherently secure and protected information channels. This approach serves as a step toward building robust quantum interconnects and advancing distributed quantum information technology mediated by topology.

The continuous increase of quantum computing power depends on reliable quantum optical interconnects to orchestrate distributed quantum systems [1,2]. However, a significant challenge is the intrinsic vulnerability of quantum states to noise and environmental disturbances. While pre-established entanglement between distant nodes could, in theory, allow for lossless state transfer [3-6], quantum states rapidly degrade with even slight loss of coherence, causing distortion of quantum information. Although quantum error correction and entanglement purification algorithms can help mitigate noise and restore quantum information [7-10], they generally require substantial resources and precise control, which limits their scalability and practical feasibility. As a result, it is highly desirable to demonstrate a strategy for loading and transferring quantum information in a robust manner, even in cases of strongly degraded quantum entanglement fidelity.

Mathematically, topology describes the properties of a structure that remain unchanged under continuous deformation. Typically characterized by topological invariants, this fundamental characteristic offers topological protection, featuring inherent robustness against perturbations. Due to its full-wave nature, light provides an ideal platform for structuring its spatial distribution and polarization state into topological forms, effectively encoding topology into photons. For example, photons can be spatially manipulated to sweep through all possible polarization states multiple times, creating optical skyrmions [11-15]. The corresponding wrapping number, characterizing how many times the spin texture of the wavefunction wraps around the spin Poincaré sphere, is the topological invariant, here a skyrmion number. When information is directly encoded on optical skyrmions, even though the local polarization of photons may fluctuate in a non-ideal environment, the total number of these sweeps (i.e., the skyrmion number) remains constant, yielding topologically protected information processing. The robustness persists even in the presence of scattering, turbulence, or other environmental noise [16,17]. Remarkably, topological protection is being expanded beyond classical light to quantum entangled photons, where the mathematical structure of skyrmions is manifested by mapping the polarization of one photon onto the quantum-correlated wavefront of the other, creating non-local skyrmions [18-20]. Similar to their classical counterpart, quantum optical skyrmions are expected to enable topologically protected quantum information processing, facilitating resilient distributed quantum technologies. Achieving this requires locally encoding the skyrmion topology on a single photon and then transmitting the topology through entanglement distribution. In this work, we integrate the generation and control of skyrmion topological textures directly into on-chip quantum photonic architectures, where integrated photonics offers inherent advantages in scalability, stability, and reconfigurability over bulk optics. The high precision and fine-tuning by on-chip photonic control make it well-suited for manipulating topological features on entangled photon states, yielding entanglement-assisted topology loading and transfer of quantum optical skyrmions.

Figure 1 shows our on-chip scheme to load the local skyrmion topology in the spin texture of a single photon $|\varphi\rangle_L = \frac{1}{\sqrt{2}}(|0, L\rangle_A + |l, R\rangle_A)$ and transfer its skyrmion

topology to a nonlocal entangled state via the following mapping:

$$|\varphi\rangle_L \to |\psi\rangle_{NL} = \frac{1}{\sqrt{2}}(|0\rangle_A |H\rangle_B + |l\rangle_A |V\rangle_B), \quad (1)$$

where subscripts A and B denote two correlated photons, 0 and $l$ are two orbital angular momentum (OAM) topological charges, and L, R, H, and V represent left-hand circular (spin-up), right-hand circular (spin-down), horizontal and vertical polarizations, respectively. Here, the local skyrmion state $|\varphi\rangle_L$ is loaded using an integrated spin-orbit locked microring resonator, which features two independent sets of the angular gratings inscribed at the inner and outer sidewalls with $N_{in}$ and $N_{out}$ scatterers, respectively. Specifically, photon A is coupled into a clockwise (CW) whispering gallery mode (WGM) inside the resonator with an azimuthal quantum number of $M = N_{in} + 1$. The WGM is subsequently extracted into the free-space Bessel mode by both sets of angular gratings. Through the spin-orbit locking mechanism and phase-matching conditions, the inner angular grating generates a spin-up polarized mode with an OAM charge of $0 = M - N_{in} - 1$, while the outer angular grating delivers a spin-down component with an OAM charge of $l = M - N_{out} + 1$ [14,15,21-23]. Their superposition forms a non-separable state of $|\varphi\rangle_L$. This state

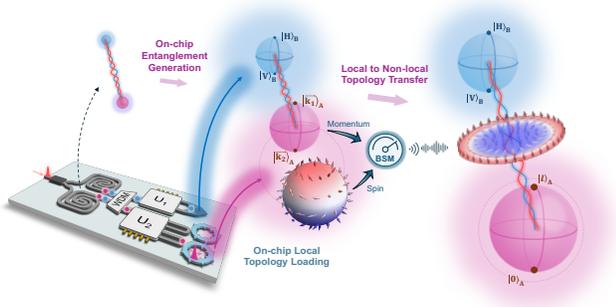

FIG. 1. Schematic of entanglement-assisted topology loading and transfer. The photonic integrated circuit, consisting of the generation of bi-photon path entanglement, wavelength division multiplexing, and unitary operations ($U_1$ and $U_2$), is used to generate entangled A (purple) and B (blue) in the state $|\chi\rangle_{AB} = \frac{1}{\sqrt{2}}\left(|\vec{k_1}\rangle_A |H\rangle_B + |\vec{k_2}\rangle_A |V\rangle_B\right)$. Simultaneously, on-chip spin-orbit-locked micro-ring resonators are applied to locally encode the topological structure of a skyrmion $|\varphi\rangle_L = \frac{1}{\sqrt{2}}(|0,L\rangle_A + |l,R\rangle_A)$ onto photon A, i.e., on-chip local topology loading. By leveraging inherent entanglement between photons A and B, performing a spin-momentum Bell state measurement (BSM) on photon A enables topology transfer from the local skyrmion structure associated with photon A to the non-local skyrmion structure shared between photons A and B, as manifested by their corresponding Stokes field distributions, respectively.

is loaded with local topology characterized by the skyrmion number $N_{sk}$.

To achieve the topology transfer illustrated in Eq. (1), a bi-photon path entanglement state is initially generated: $|\Psi_1\rangle_{AB} = \frac{1}{\sqrt{2}}(|P_1\rangle_A |P_1\rangle_B + |P_2\rangle_A |P_2\rangle_B)$, where $P_1$ and $P_2$ represent two distinct on-chip paths. This process is facilitated by pumping two on-chip spiral waveguides with a femtosecond laser pulse at frequency $\omega_{pump}$, which probabilistically drives spontaneous four-wave mixing (SFWM) to generate photon pairs at frequencies $\omega_A$ and $\omega_B$ [24,25]. The nonlinear processes in the two spirals are coherently superposed, generating the state $|\Psi_1\rangle_{AB}$ upon repacking the two photons at different frequency through the unbalanced Mach-Zehnder interferometers (MZIs) [25]. Photon B is then directed into a two-dimensional (2D) grating coupler, which coherently converts $|P_1\rangle_B$ and $|P_2\rangle_B$ paths into free-space modes carrying H and V polarizations, respectively [26,27]. Simultaneously, photon A, propagating along the $|P_1\rangle_A$ and $|P_2\rangle_A$ paths, is routed into two identical microring resonators, where the local topological state $|\varphi\rangle_L$ is coherently encoded onto the photon. Upon extraction from two distinct resonators and collection by an objective lens, photon A acquires two different momenta, $\vec{k_1}$ and $\vec{k_2}$. This process produces a tensor product state that intertwines the local skyrmion with inherent entanglement between two photons:

$$|\Psi_2\rangle_{AB} = |\varphi\rangle_L \otimes \frac{1}{\sqrt{2}}\left(|\vec{k_1}\rangle_A |H\rangle_B + |\vec{k_2}\rangle_A |V\rangle_B\right) = |\varphi\rangle_L \otimes |\chi\rangle_{AB}. \quad (2)$$

Analogous to the entanglement swapping process [28,29], the state can be decomposed using the spin-momentum Bell states of photon A as follows:

$$|\Psi_2\rangle_{AB} = \frac{1}{2}(|\Phi^+\rangle_A |\psi\rangle_{NL} + |\Phi^-\rangle_A \sigma_{z,B} |\psi\rangle_{NL} + |\Psi^+\rangle_A \sigma_{x,B} |\psi\rangle_{NL} - i|\Psi^-\rangle_A \sigma_{y,B} |\psi\rangle_{NL}), \quad (3)$$

where $|\Phi^\pm\rangle_A = \frac{1}{\sqrt{2}}(|L,\vec{k_1}\rangle_A \pm |R,\vec{k_2}\rangle_A)$ and $|\Psi^\pm\rangle_A = \frac{1}{\sqrt{2}}(|L,\vec{k_2}\rangle_A \pm |R,\vec{k_1}\rangle_A)$ represent the four spin-momentum Bell basis states, and the operators $\sigma_{x,B}$, $\sigma_{y,B}$, and $\sigma_{z,B}$ denote the Pauli matrices acting on the polarization degree of freedom of photon B. In this scenario, performing a spin-momentum Bell state measurement (BSM) on photon A and applying the corresponding Pauli operations based on the BSM outcomes effectively erase the spin and momentum information in photon A. In doing so, the process establishes a hybrid entangled state between photons A and photon B, which transfers the local topology encoded in $|\varphi\rangle_L$ to the non-local shared topology in $|\psi\rangle_{NL}$, realizing the mapping function described in Eq. (1).

The photonic integrated circuit for encoding local topology and distributing bi-photon entanglement is fabricated on a $Si_3N_4$-on-$SiO_2$ platform with a 300 nm-thick $Si_3N_4$ layer (Fig. 2(a)). The spiral waveguides are tailored with a width of 650 nm to optimize momentum matching in the SFWM process, thereby maximizing the efficiency of the correlated photon generation. Additionally, the microring resonators are designed with a width of 600 nm to enhance spin purity at the edge, facilitating the spin-orbit locking. We first demonstrate the loading of local topology through resonators with $N_{in} = N_{out} = M - 1$, resulting in a local skyrmion state of $\frac{1}{\sqrt{2}}(|0, L\rangle_A + |2, R\rangle_A)$ with a skyrmion number $N_{sk} = 2$. In experiments, the local topology of the single photon skyrmion is characterized by performing quantum state tomography [30] on both the spin and OAM of photon A extracted from the microring resonators. The reconstructed density matrix $\rho_L = |\varphi\rangle_L\langle\varphi|_L$ shows a fidelity of 92.6±1.1% and the Stokes field distribution at the wavefront is subsequently derived from the expectation values of $\rho_L$ (Fig. 2(b)), yielding an experimentally retrieved local skyrmion number $N_{sk} = 1.99586 \pm 0.00006$ in agreement with theoretical expectations (see Supplementary Material [31]). The bi-photon entanglement $|\chi\rangle_{AB}$ is characterized by performing quantum state tomography on the momentum of photon A, extracted from the micro-ring resonators, and the polarization of photon B, captured from the 2D grating coupler (see Supplementary Material [31]). The experimentally retrieved density matrix $\rho_\chi = |\chi\rangle_{AB}\langle\chi|_{AB}$ exhibits a high quantum fidelity of 95.4±1.1% (Fig. 2(c)). The topology transfer of the skyrmion structure is achieved by transitioning from the local skyrmion on photon A to the non-local correlation between entangled photons A and B. This is done by first performing the spin-momentum BSM on photon A. A momentum-dependent spin flip operation is designed to resolve the non-separability of the spin-momentum Bell states, facilitating efficient BSM with independent operations on spin and momentum (see Supplementary Material [31]). Consequently, the spin-momentum correlation of photon A is projected onto states $|\Phi^\pm\rangle_A$ and $|\Psi^\pm\rangle_A$ as described in Eq. (3), which generates the non-local skyrmion states $|\psi\rangle_{NL}$, $\sigma_{z,B}|\psi\rangle_{NL}$, $\sigma_{x,B}|\psi\rangle_{NL}$, and $-i\sigma_{y,B}|\psi\rangle_{NL}$, respectively. Quantum state tomography is then conducted on the tensor-product Hilbert space formed by the OAM of photon A and the polarization of photon B, yielding the four density matrices with an averaged fidelity of 94.1±1.2% (Fig. 2(d)) (see Supplementary Material [31]). The quantum Stokes parameter distributions corresponding to the four density matrices are calculated after applying the relevant Pauli operations, featuring non-local skyrmion numbers of 1.99558±0.00004, 1.99573±0.00004, 1.99532±0.00006, and 1.99564±0.00010, respectively. This convincingly validates our scheme of topology loading and transfer through entanglement distribution.

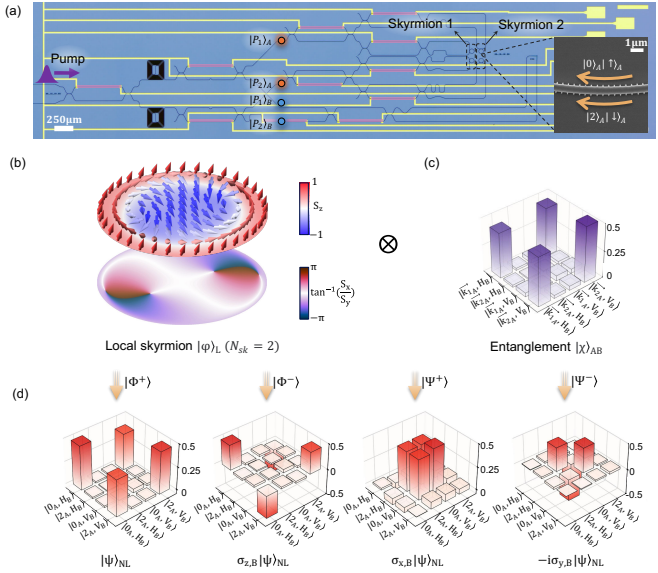

FIG. 2. Experimental characterization of the topology loading and transfer. (a) Optical microscope image of the photonic integrated circuits used to implement the scheme in Fig. 1. A femtosecond pump pulse is split evenly to pump two spiral waveguides, generating the path-entangled photon pair $|\Psi_1\rangle_{AB} = \frac{1}{\sqrt{2}}(|P_1\rangle_A|P_1\rangle_B + |P_2\rangle_A|P_2\rangle_B)$ via spontaneous four-wave mixing (SFWM). Mach-Zehnder interferometer (MZI) arrays then route entangled photons A and B into two identical microring resonators and a two-dimensional grating coupler, respectively, creating the local skyrmion and the entanglement state $|\Psi_2\rangle_{AB} = |\varphi\rangle_L \otimes |\chi\rangle_{AB}$. Four microrings are fabricated, with the left two generating the skyrmion-1 state and the right two generating the skyrmion-2 state. Results for the skyrmion-2 state are presented in the manuscript. The inset shows a scanning electron microscope image of the skyrmion-2 microring resonator with angular gratings on its inner and outer sidewalls, generating $|0, L\rangle_A$ and $|2, R\rangle_A$ for clockwise whispering gallery modes. (b) Measured Stokes parameters distribution for the local skyrmion state. The color of the upper plane shows the $S_z$ component, while the lower plane illustrates the phase of the Stokes vector in the $S_x$, $S_y$ plane, showing a phase winding of $4\pi$ and corresponding to a skyrmion number $N_{sk} = 2$. (c) Retrieved density matrix for the entanglement state $|\chi\rangle_{AB}$, exhibiting a quantum fidelity of 95.4±1.1%. (d) Retrieved density matrices for four non-local skyrmion states after topology transfer: $|\psi\rangle_{NL}$, $\sigma_{z,B}|\psi\rangle_{NL}$, $\sigma_{x,B}|\psi\rangle_{NL}$, and $-i\sigma_{y,B}|\psi\rangle_{NL}$. Each matrix corresponds to one of the four spin-momentum Bell state measurement outcomes ($|\Phi^\pm\rangle_A$ and $|\Psi^\pm\rangle_A$), with quantum fidelity of 93.5±1.4%, 94.0±1.4%, 94.5±1.0%, and 94.5±1.1%, and non-local skyrmion numbers of 1.99558±0.00004, 1.99573±0.00004, 1.99532±0.00006, and 1.99564±0.00010, respectively. The uncertainties were determined using a Monte Carlo method that considers Poissonian photon statistics [35].

The virtue of topological protection is the robust preservation of the topological invariant throughout the topology transfer process, even in the presence of high environmental noise and largely degraded quantum correlations which are typical challenges in distributed quantum systems. Specifically, in our case, the winding nature of the skyrmion structure ensures that the topology remains if the quantum correlation of the entanglement state $\rho_\chi$ used to transfer the topology does not vanish (see Supplementary Material [31]). In experiments, the robustness of the topology transfer process is verified under varying levels of background and isotropic white noise, with key metrics including state fidelity $F$, purity $\gamma$, and skyrmion number $N_{sk}$ being measured (Fig. 3(a)). With the increase of background noise, both $\gamma$ and $F$ of the non-local quantum skyrmion decrease, while $N_{sk}$ remains invariant. To further evaluate topological robustness, external isotropic noise is intentionally introduced into the entanglement correlation using an incoherent white light source with varying intensities [19,36]. This noise maps the pure nonlocal skyrmion state $\rho_{NL} = |\psi\rangle_{NL}\langle\psi|_{NL}$ onto a new partially mixed density matrix $\rho'_{NL} = (1-\xi_0)\rho_{NL} + \frac{\xi_0}{4}I_{4\times4}$, scaling the degree of mixture via the parameter $\xi_0$, where $I_{4\times4}$ is the four-dimensional identity matrix characterizing the isotropic noise (see Supplementary Material [31]). As isotropic noise increases, both $\gamma$ and $F$ of the non-local quantum skyrmion continue to drop. However, $N_{sk}$ remains unchanged until the noise strength exceeds a critical threshold of $\gamma \sim 0.25$, where isotropic noise starts to dominate, overwhelming both the original state and anisotropic noise contributions. At this point, both the noisy entanglement state $\rho'_\chi$ and the nonlocal skyrmion state $\rho'_{NL}$ suffer complete degradation of quantum correlations, collapsing into the maximally mixed state with vanishing quantum discord due to the strong noise, as predicted by our theoretical analysis (see Supplementary Material [31]). This finding is further supported by the experimentally retrieved density matrices and reconstructed quantum Stokes field along the noise-increasing trajectory (Fig. 3(b)). The results demonstrate the robustness of the topology transfer process under isotropic noise until the quantum correlation is destroyed.

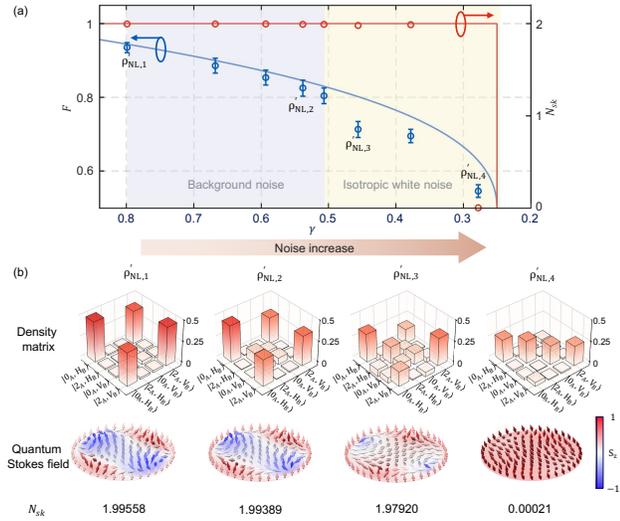

FIG. 3. Robustness of topology transfer under background and isotropic white noise. (a) The experimentally retrieved fidelity ($F$) and purity ($\gamma$) of the non-local skyrmion state after topology transfer (blue points) decrease continuously with increasing measurement background noise (purple region) and isotropic white noise (yellow region). In contrast, $N_{sk}$ (red points) remains invariant until the state reaches $\rho'_{NL,4}$, where isotropic white noise overwhelms both the original state and anisotropic noise contributions, yielding the maximally mixed state (see Supplementary Material [31]). This behavior aligns well with theoretical predictions, where the fidelity is given by $F = \sqrt{\frac{\sqrt{3(4\gamma-1)}+1}{4}}$ (blue line) (see Supplementary Material [31]), and $N_{sk}$ is preserved until $\gamma$ reaches 0.25 (red line). The error bars for $N_{sk}$ are so small that they cannot be resolved in the plot. (b) Retrieved density matrices and quantum stokes fields for four representative states, with skyrmion numbers of 1.99558±0.00004, 1.99389±0.00018, 1.97920±0.00495, and 0.00021±0.00003, respectively.

The resilience of the topology transfer is also tested by directly applying anisotropic noise to the generation of on-chip entanglement [20,32]. The on-chip phase manipulation over Photon B can facilitate a phase-flip noise channel ($\sigma_{z,B}$), leading to the depolarization of photon B, which corresponds to the contraction of the polarization Poincaré sphere into an ellipsoid, with its major axis aligned along the $S_z$ direction (Fig. 4(a)). In this scenario, a $\pi$-phase shift translates into a phase-flip error on photon B's polarization. To mimic the noise channel with controlled strengths, the measured coincidence counts are numerically mixed with the original counts without the phase shift at a ratio of $p_3$, represented as an incoherent superposition of the noiseless and noisy states in the retrieved density matrix. The reconstructed density matrices for both $\rho'_\chi$ and $\rho'_{NL}$ exhibit a continuous reduction in the off-diagonal components as $p_3$ increases, manifesting the degradation of entanglement correlations (see, for example, at $p_3 = 0.3$ shown in Fig. 4(b)). Despite the decrease of entanglement fidelity with $p_3$, the topological number of the non-local quantum skyrmion number remains intact up to a critical threshold of $p_3 = 0.5$, where the quantum discord of $\rho'_\chi$ vanishes. (Fig. 4(c)) (see Supplementary Material [31]). Similarly, a bit-phase-flip noise channel ($\sigma_{y,B}$) can be also applied, where the $|P_1\rangle_B$ and $|P_2\rangle_B$ paths for photon B need to be swapped by controlling the on-chip MZI. In this case, the contracted polarization Poincaré ellipsoid of Photon B aligns with its major axis in the $S_y$ direction (Fig. 4(d)). With the increase of a controlled ratio of $p_2$, despite the degradation of entanglement correlations (Fig. 4(e)), the topological number of the non-local quantum skyrmion number remains preserved until the critical threshold of $p_2 = 0.5$ (Fig. 4(f)) (see Supplementary Material [31]). The results confirm again that, the entanglement-assisted topology transfer process remains topologically robust against a large variety of noise and external perturbations.

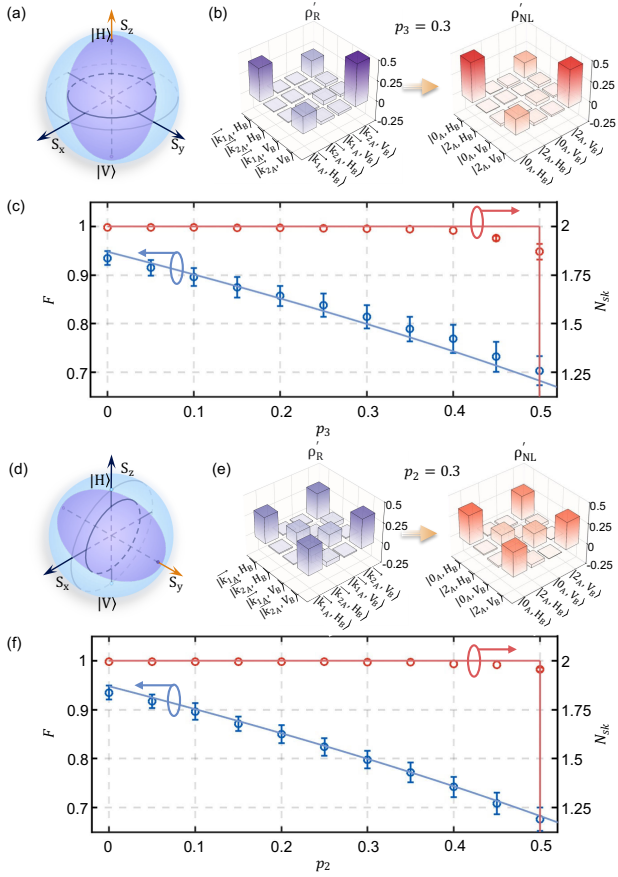

FIG. 4. Robustness of topology transfer under anisotropic noise. (a) and (d) Pictorial representation of photon B's Poincaré sphere contraction into an ellipsoid under the phase-flip ($\sigma_{z,B}$) and bit-phase-flip ($\sigma_{y,B}$) noisy channels, with the major axis aligned along the $S_z$- and $S_y$- directions, respectively. (b) and (e) Retrieved density matrices for the entanglement state $\rho'_\chi$ after the on-chip generation and the non-local skyrmion state $\rho'_{NL}$ with noise ratios $p_3 = 0.3$ and $p_2 = 0.3$, respectively. (c) and (f) Scatter points represent the simulated results with experimental data, where the fidelity of the non-local skyrmion state (blue points) decreases continuously with increasing noise ratios ($p_3$ for phase-flip in c and $p_2$ for bit-phase-flip in f), while $N_{sk}$ (red points) remains invariant until the noise ratio approaches 0.5. These observations align with theoretical predictions (solid lines), where fidelity $F$ decreases with noise ratio $p$ according to $F = \sqrt{1 - \frac{3}{4}\lambda_1 - (1-\lambda_1)p}$ (blue lines) with $\lambda_1$ characterizing the fidelity of local skyrmion (see Supplementary Material [31]), while $N_{sk}$ remains preserved until the noise ratio reaches 0.5 (red lines). The error bars for some of the $N_{sk}$ values are so small that they cannot be resolved in the plot.

We have demonstrated the integrated loading of local topological features onto single photons and their robust transfer to non-local quantum states through a swapping protocol, enabled by chip-generated quantum entanglement. This enables a seamless transfer of the topological features regardless of the quality of the resources used in the generation, state transfer, and measurement procedures. Remarkably, our further analysis reveals that the transferred topological invariant persists even when entanglement—as quantified by concurrence—is entirely lost, yet it fails to survive once the quantum discord of the link vanishes. This finding suggests that topological protection might be related to a form of quantum correlation that lies between entanglement and discord [37] (see Supplementary Material [31]). Therefore, our approach exhibits exceptionally robust skyrmion loading and transfer from local to non-local states, with their topological invariant well preserved even in the presence of large environmental noise and significant entanglement degradation, if the quantum correlations are not fully lost. The demonstrated topology transfer process is inherently compatible with quantum teleportation protocols when ancillary photons are employed [3,4], providing noise-resilient quantum links to connect independent quantum systems in a distributed manner while using topological encoding as a resource. From a foundational perspective, our work explores how global topological features can be loaded, transferred and preserved through photon entanglement, pointing toward a new class of quantum communication protocols where resilience is built not only through redundancy and correction, but through the structure of the quantum states themselves. Topology-assisted quantum information is still in its very early stage, but various mechanisms are being explored to encode and decode information onto the topological invariants associated with the spin or/and wavefunction texture of photons [11,38]. Central to our demonstration is precise and scalable control of photon entanglement and their flexible routing and switching on-chip, making integrated photonics an ideal platform for conducting complex topological protocols. Alongside these exciting developments, the powerful integrated toolbox presented in our work offers a promising foundation to inspire further efforts in leveraging topology to realize robust and reliable quantum communication interconnects for achieving distributed quantum systems.

*yijie.shen@ntu.edu.sg
†andrew.forbes@wits.ac.za
‡fenglia@seas.upenn.edu